\documentclass{article}
\usepackage{spconf,amsmath,graphicx,hyperref}
\usepackage{threeparttable}
\usepackage{booktabs}
\usepackage{listings}
\usepackage{listingsutf8}
\usepackage{tabularx}

\title{Jamendo-QA: A Large-Scale Music Question Answering Dataset}
\name{
Junyoung Koh$^{1,2,3\dagger}$\thanks{$^{\dagger}$Equal contribution. Corresponding author. This research was supported by Brian Impact Foundation, a non-profit organization dedicated to the advancement of science and technology for all.}, 
Soo Yong Kim$^{2,4,*}$\thanks{$^{*}$Equal contribution.}, 
Yongwon Choi$^{2,*}$, 
Gyu Hyeong Choi$^{2,5}$
}
\address{
$^{1}$Department of Artificial Intelligence, Yonsei University \\
$^{2}$MAAP LAB, MODULABS \quad $^{3}$KRAFTON \quad $^{4}$AI Matics \\ $^{5}$Department of Media Software, Sungkyul University \\
\texttt{solbon1212@yonsei.ac.kr}
}
\begin{document}
%
\maketitle
\begin{abstract}
We introduce \textbf{Jamendo-QA}, a large-scale dataset for Music Question Answering (Music-QA). 
The dataset is built on freely licensed tracks from the Jamendo platform and is automatically annotated using the Qwen-Omni model. 
Jamendo-QA provides question-answer pairs and captions aligned with music audio, enabling both supervised training and zero-shot evaluation. 
Our resource aims to fill the gap of music-specific QA datasets and foster further research in music understanding, retrieval, and generative applications. In addition to its scale, Jamendo-QA covers a diverse range of genres, instruments, and metadata attributes, allowing robust model benchmarking across varied musical contexts. 
We also provide detailed dataset statistics and highlight potential biases such as genre and gender imbalance to guide fair evaluation. 
We position Jamendo-QA as a scalable and publicly available benchmark that can facilitate future research in music understanding, multimodal modeling, and fair evaluation of music-oriented QA systems.
\end{abstract}
\begin{keywords}
Music Question Answering, Music QA Dataset, Generative AI, Music Understanding, Music Information Retrieval
\end{keywords}
\section{Introduction}
\label{sec:intro}

\begin{figure}[t]
    \centering
    \includegraphics[width=0.8\linewidth]{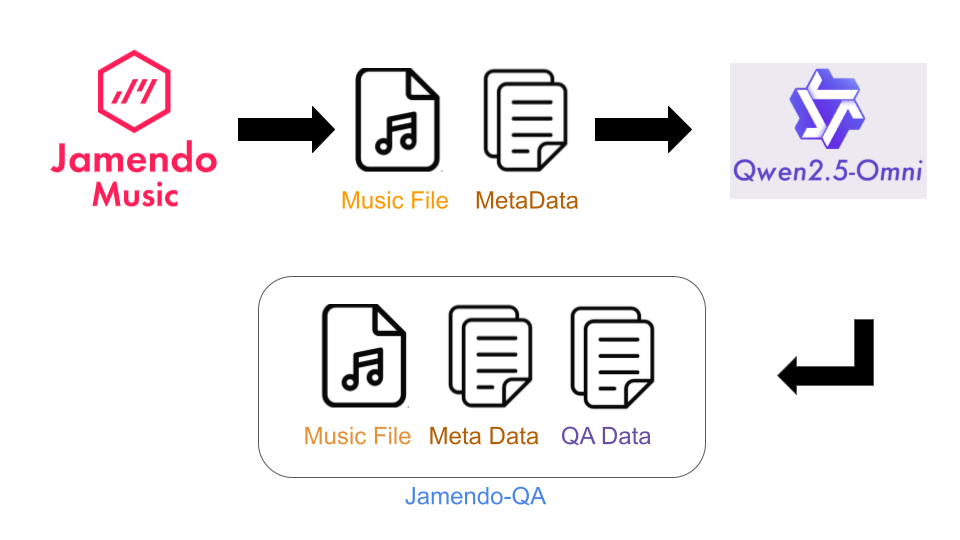}
    \caption{The proposed pipeline for automatic generation of the Jamendo-QA dataset. The process utilizes the Qwen-Omni Multimodal LLM to generate captions and question-answer pairs from raw music audio and associated metadata.}
    \label{fig:architecture}
\end{figure}

\begin{table*}[t]
  \caption{Comparison of related datasets}
  \label{tab:musicqa-comparison}
  \centering
  \small
  \setlength{\tabcolsep}{1pt} 
  \renewcommand{\arraystretch}{1.1} 
  \begin{tabularx}{\linewidth}{l c c c X}
    \toprule
    \textbf{Name} & \textbf{Count} & \textbf{Task} & \textbf{Domain} & \textbf{Method (how created)} \\
    \midrule
    MUSIC-AVQA \cite{musicavqa} & 45K  & QA      & Music (Audio+Video) & Music video clips with QA annotations; audio–visual QA pairs curated for musical reasoning. \\
    MusicQA \cite{mullama} & 13K  & QA      & Music (Audio)       & Multimodal LLM for music/audio QA; MERT encoder + adapter (linear+SiLU \cite{silu}). \\
    MTG-Jamendo \cite{mtgjamendo}   & 55K  & Tagging & Music (Audio)       & Jamendo tracks with crowd-sourced multi-label tags (genre/instrument/mood). \\
    JamendoMaxCaps \cite{jamendomaxcaps} & 360K & Captioning & Music (Audio) & Large-scale captions over Jamendo; auto/semi-auto from metadata/tags + Qwen2-Audio \cite{qwen2audio}. \\
    LP-MusicCaps \cite{lpmusiccaps} & 542K & Captioning & Music (Audio)    & Language-paired music captions at scale; pretrained encoders + GPT-3.5 turbo \cite{gpt3.5turbo}. \\
    MusicXQA \cite{musicxqa}        & 1.29M & QA      & Music (MIDI+sheet)  & QA pairs derived from symbolic music data (MIDI and sheet music) using MLLM-based generation. \\
    \midrule
    \textbf{Jamendo-QA (ours)} & \textbf{37K} & \textbf{QA} & \textbf{Music (Audio)} & \textbf{QA pairs and captions auto-generated on Jamendo using Qwen-Omni.} \\
    \bottomrule
  \end{tabularx}
\end{table*}

\begin{table}[t]
\centering
\caption{Statistical summary of the dataset}
\label{tab:audio-stats}
\small
\renewcommand{\arraystretch}{1.1}
\begin{tabularx}{\linewidth}{l *{3}{>{\centering\arraybackslash}X}}
\toprule
\textbf{Measure} & \textbf{Mean} & \textbf{Median} & \textbf{Mode} \\
\midrule
Length (sec)      & 233.4 & 220.5 & 240.0 \\
SNR (dB)          & 12.48 & 11.69 & 9.38 \\
RMS Energy        & 0.195 & 0.191 & 0.120 \\
Zero Crossing Rate& 0.057 & 0.054 & 0.038 \\
\bottomrule
\end{tabularx}
\end{table}

Music Question Answering (Music-QA) focuses on answering natural language questions about music audio. Unlike general audio tagging \cite{htsat} or captioning \cite{lpmusiccaps, koizumi2020audiocaptioningusingpretrained, clotho}, Music-QA requires fine-grained reasoning over temporal and spectral structures in music, often combining semantic, structural, and stylistic knowledge. However, the lack of large-scale and diverse QA datasets has limited the exploration of this field. Existing works mainly rely on small, manually curated datasets or focus on broader audio domains rather than music-specific QA. Recently, contrastive audio-language pretraining models such as CLAP \cite{elizalde2023clap} have shown strong generalization across audio tasks, but they are not directly designed for music-focused QA.

Recent research \cite{musicxqa} has begun to address this challenge by proposing new datasets and models for Music-QA. For example, MU-LLaMA \cite{mullama} introduced an adapter-based architecture leveraging MERT \cite{mert} audio features for downstream QA tasks, while MuMu-LLaMA \cite{mumullama} extended this idea into a broader multimodal framework covering audio, text, and video, incorporating generative models such as AudioLDM2 \cite{audioldm, audioldm2} and MusicGen \cite{musicgen}. These efforts highlight the potential of large language models combined with audio encoders for music understanding, but they remain constrained by the limited availability of high-quality QA data in the music domain.

To partially overcome this, cross-domain approaches such as MUSIC-AVQA \cite{musicavqa} have been developed, extending the problem into audio-visual settings. MUSIC-AVQA pairs video and music with question-answer annotations and employs CNN and LSTM-based \cite{lstm} architectures for reasoning. While these datasets demonstrate the feasibility of QA in multimodal contexts, they do not provide music-centric QA resources that directly address purely musical audio without reliance on visual modalities.

Motivated by these gaps, we introduce the \textbf{Jamendo-QA dataset}, a large-scale QA dataset derived from the Jamendo music platform. Leveraging Qwen-Omni \cite{qwenomni} for automated question-answer and caption generation, our dataset provides a scalable, freely licensed resource for advancing research in music QA. By combining open-access music with generative QA annotations, Jamendo-QA represents one of the first attempts to construct a dedicated large-scale QA dataset for music understanding.

\begin{figure}[t]
    \centering
    \includegraphics[width=0.9\linewidth]{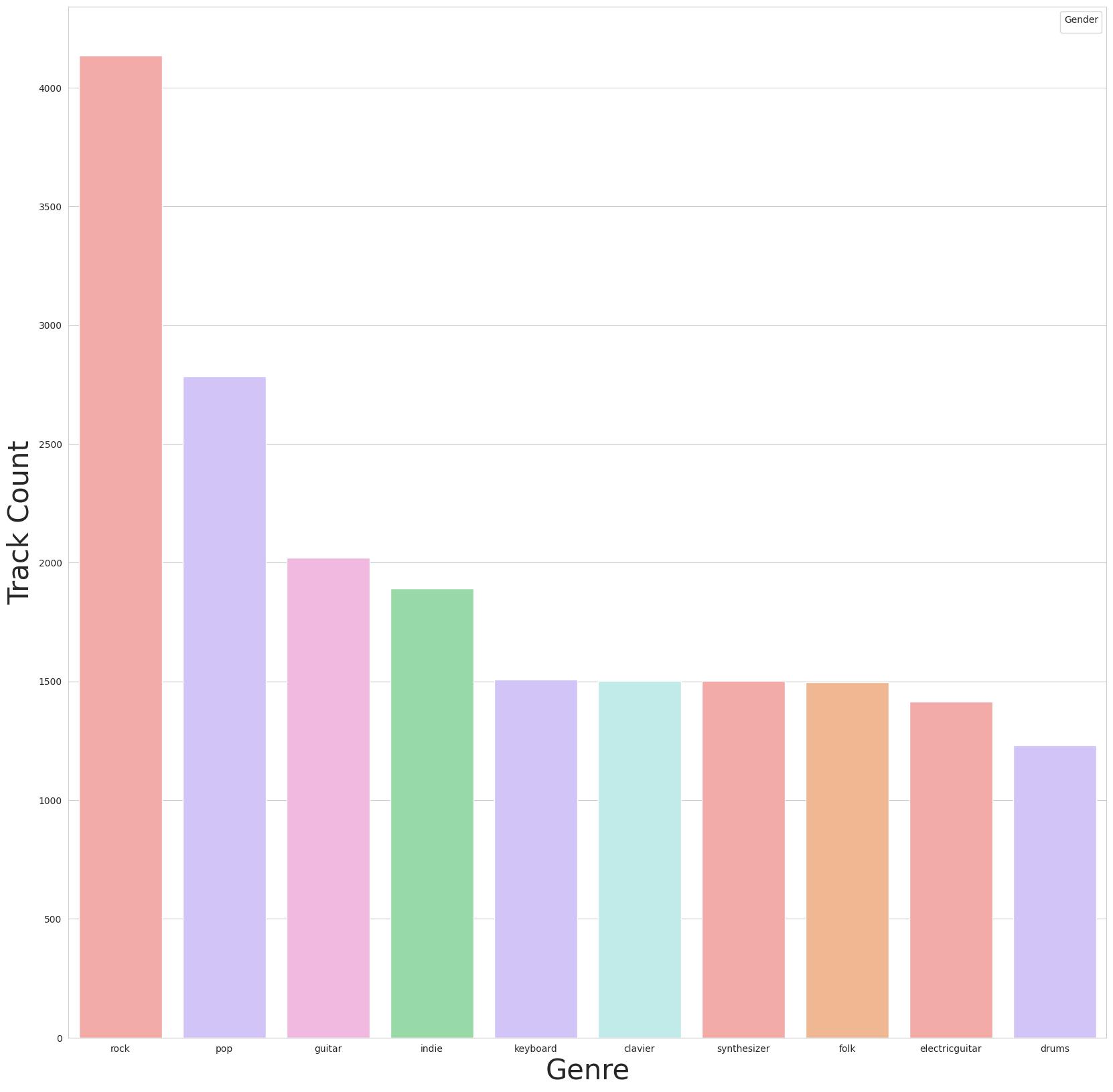}
    \caption{Track counts for the top-10 genres in Jamendo-QA.}
    \label{fig:track-counts}
\end{figure}

\section{Method}
\label{sec:method}
As illustrated in Figure~\ref{fig:architecture}, our approach leverages a multimodal large language model, Qwen-Omni, to automatically generate question-answer pairs and captions from music audio. This generative pipeline enables the creation of a large-scale dataset, Jamendo-QA, without manual annotation, thereby addressing the data scarcity issue in Music-QA.

The pipeline begins with a raw music track from the Jamendo platform. This audio input is first processed by a specialized audio encoder, which transforms the raw audio into a sequence of meaningful audio embeddings. These embeddings capture the acoustic characteristics of the music, such as timbre, rhythm, and harmony.

Simultaneously, the model can also be provided with existing metadata from the track (e.g., genre, instruments, artist info). This text metadata is processed by a text encoder to create text embeddings. Both the audio and text embeddings are then fed into the core Qwen-Omni Multimodal LLM.

Inside the LLM, the audio and text features are fused and contextualized. The model uses this combined information to generate a variety of outputs:
\begin{itemize}
    \item \textbf{Music Captions}: Descriptive text summaries of the music.
    \item \textbf{Question Generation}: A diverse set of questions about the music (e.g., "What is the dominant instrument?", "What is the mood of this song?").
    \item \textbf{Answer Generation}: Corresponding answers to the generated questions, based on the input audio and metadata.
\end{itemize}
This generative process results in the Jamendo-QA dataset, which contains high-quality, automatically generated question-answer pairs and captions aligned with the original music audio. This dataset can be used for training and evaluating models for music understanding and retrieval tasks.

\section{Analysis}
To better understand the characteristics of \textbf{Jamendo-QA}, we conduct a
three-stage analysis: (i) \textit{Univariate analysis} of key metadata 
distributions, (ii) \textit{Multivariate/correlation analysis} exploring 
relationships between attributes, and (iii) \textit{QA-level analysis} 
of conversation patterns.

\subsection{Univariate Analysis}

\begin{figure}[t]
    \centering
    \includegraphics[width=0.95\linewidth]{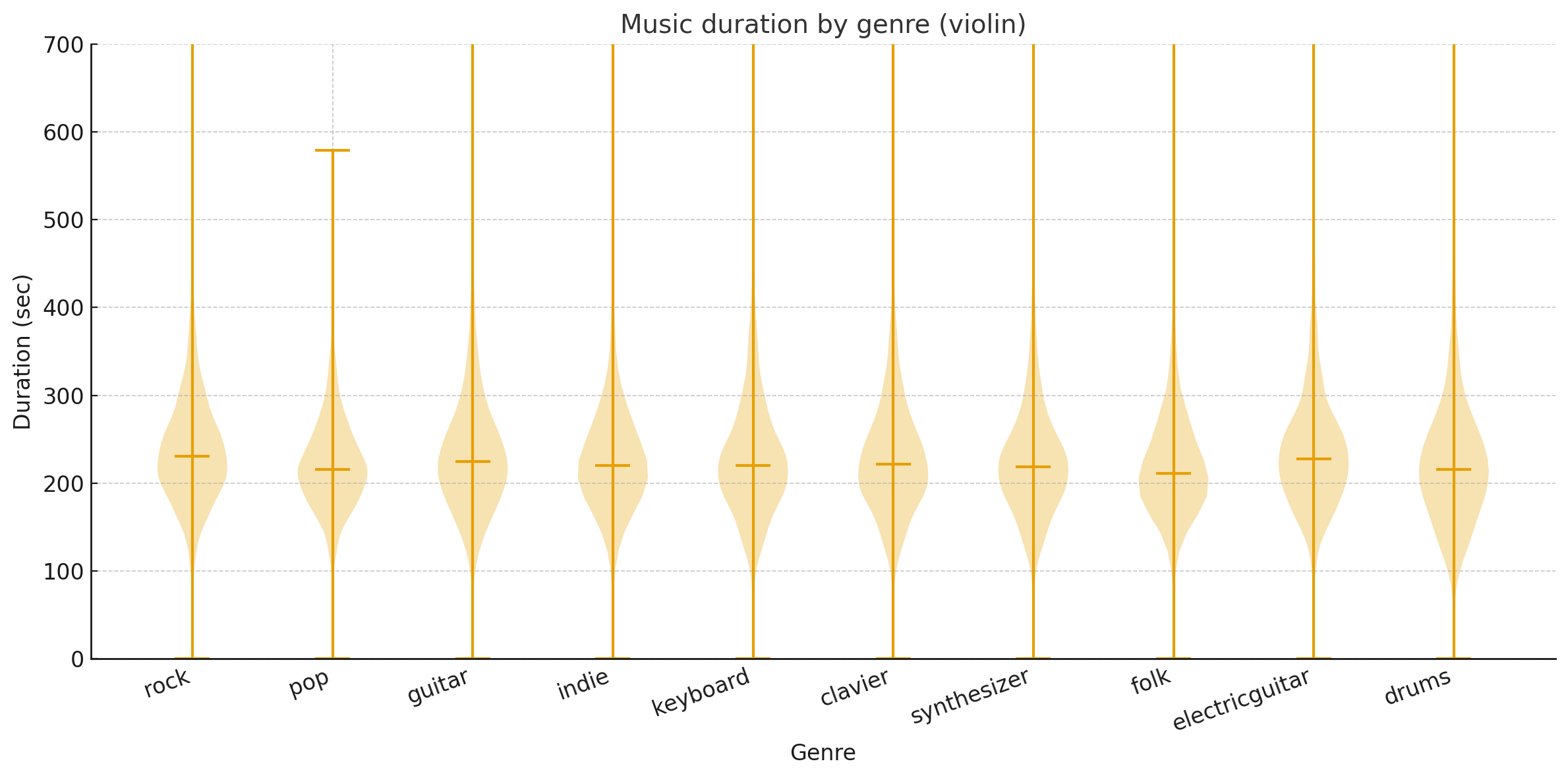}
    \caption{Violin plot of music duration by genre. Most tracks range between 200--300 seconds.}
    \label{fig:duration-dist}
\end{figure}

We first investigate global distributions of the dataset. 
Figure~\ref{fig:track-counts} shows that the dataset is dominated by 
\texttt{rock} and \texttt{pop}, followed by \texttt{guitar} and 
\texttt{indie}. Figure~\ref{fig:duration-dist} depicts the distribution of 
track durations using violin plots, indicating that most tracks cluster 
between 200--300 seconds, with a small number of outliers reaching up to 
600 seconds. These observations highlight a realistic coverage of 
standard-length songs.

\subsection{Multivariate / Correlation Analysis}

\begin{figure}[t]
    \centering
    \includegraphics[width=0.95\linewidth]{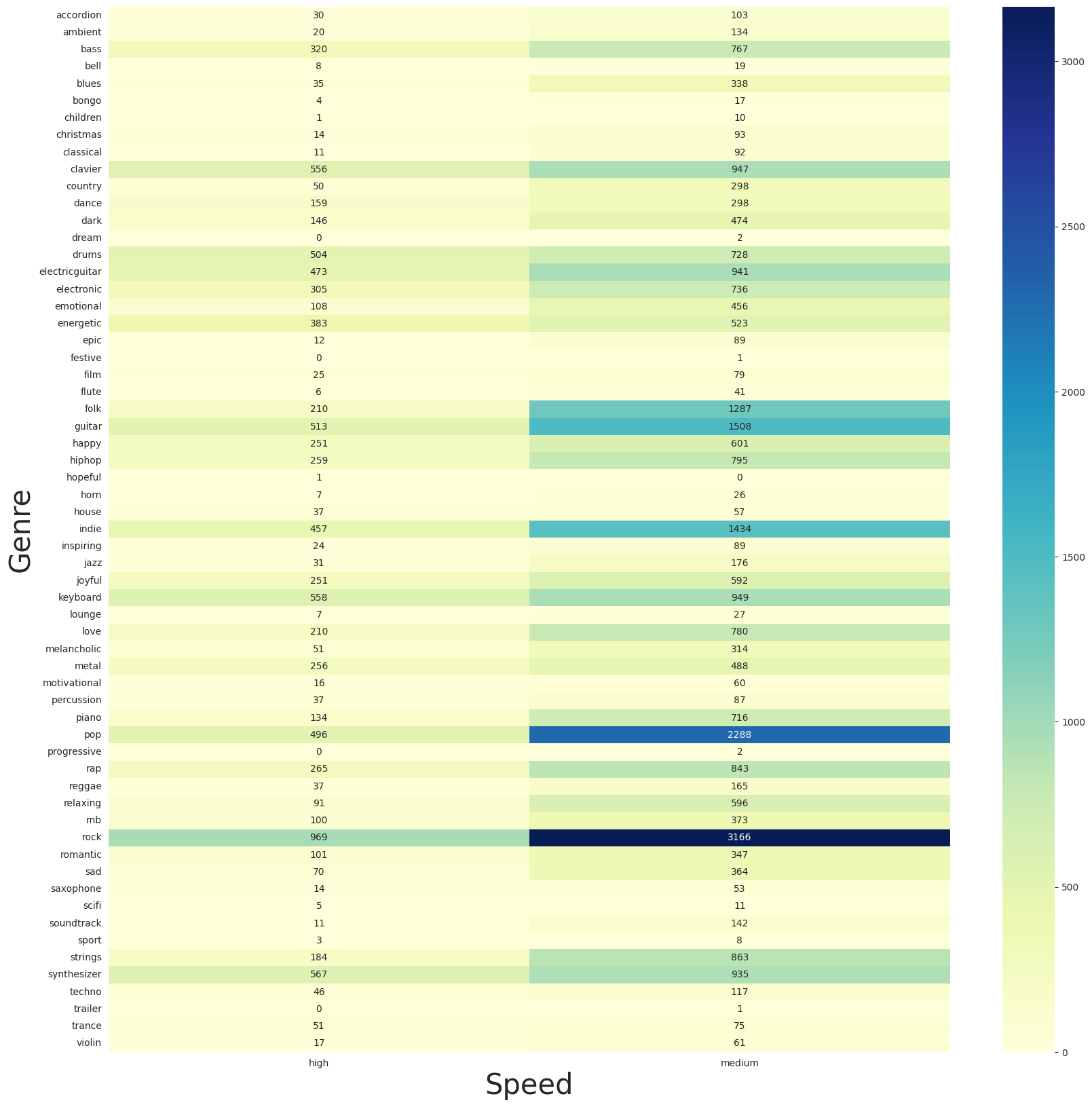}
    \caption{Relationship between genre and tempo. 
    Medium tempo is dominant across most genres, 
    with \texttt{rock} and \texttt{pop} being particularly frequent.}
    \label{fig:genre-speed}
\end{figure}

\begin{figure}[t]
    \centering
    \begin{minipage}[t]{0.48\linewidth}
        \centering
        \includegraphics[width=\linewidth]{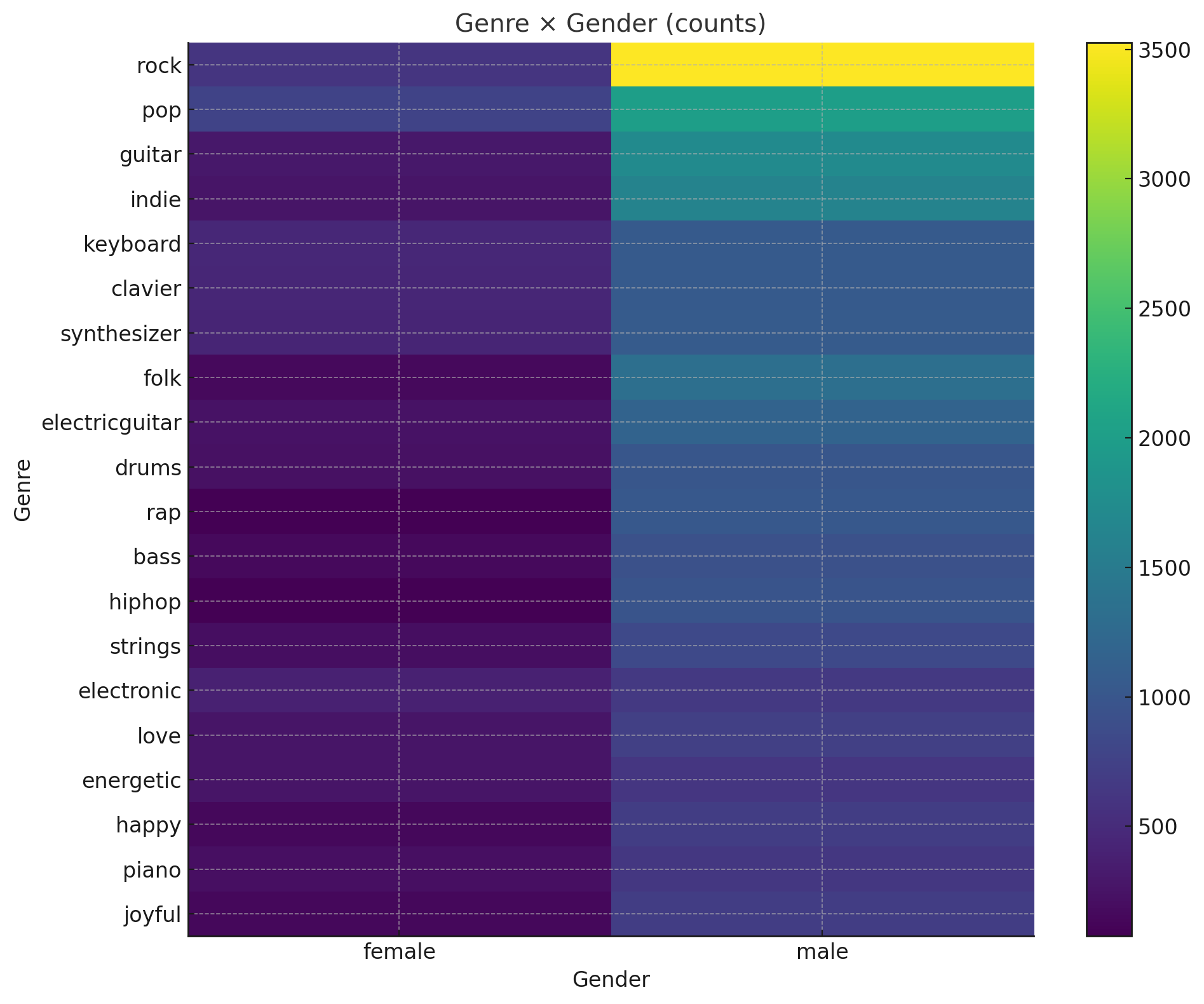}
    \end{minipage}\hfill
    \begin{minipage}[t]{0.48\linewidth}
        \centering
        \includegraphics[width=\linewidth]{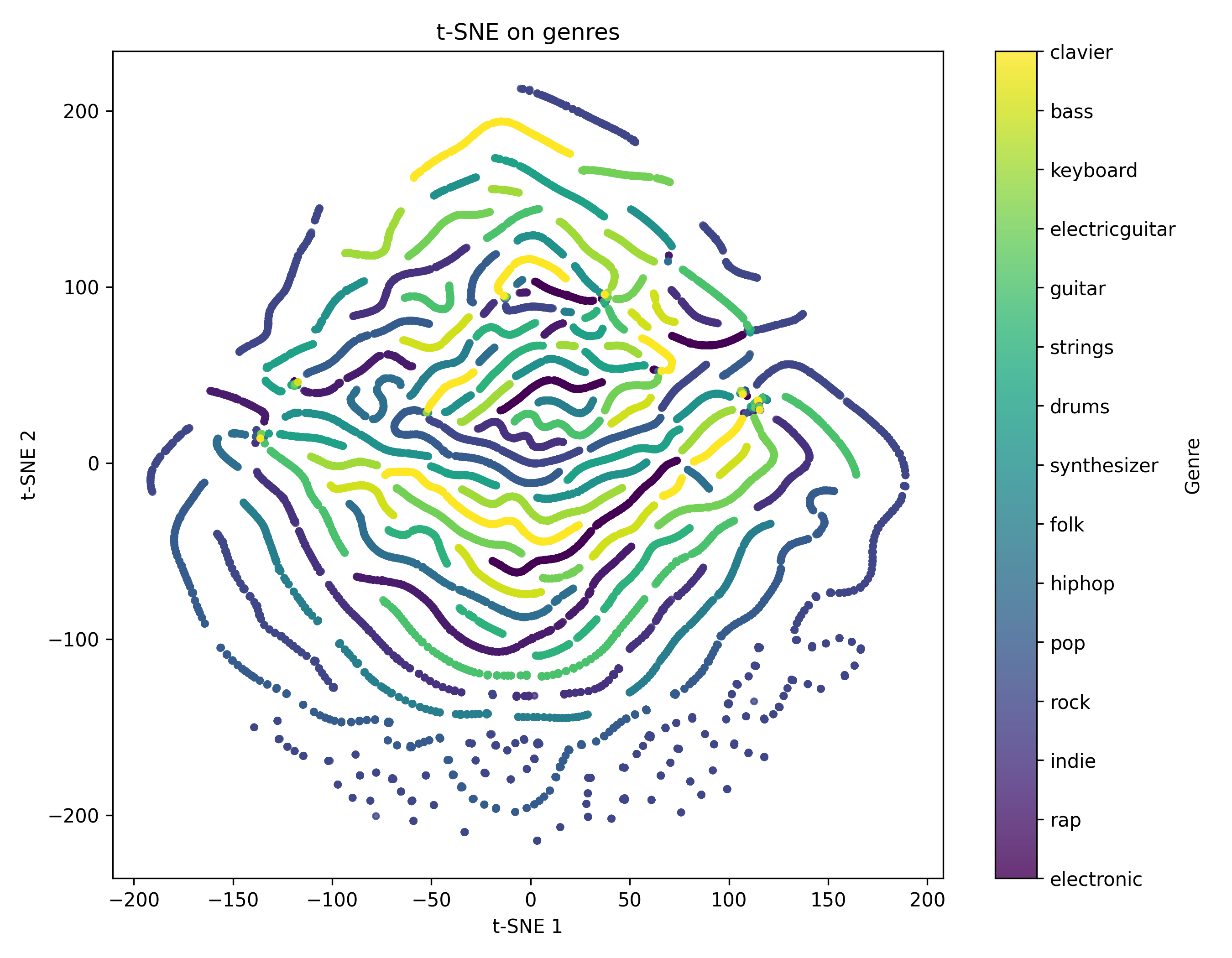}
    \end{minipage}
    \caption{Metadata visualizations. Left: artist gender distribution across genres. Right: 2D t-SNE embedding (colored by genre) showing genre separability.}
    \label{fig:metadata-viz}
\end{figure}

Next, we explore cross-feature relationships. 
Figure~\ref{fig:genre-speed} shows the co-occurrence between genre and tempo.
Figure~\ref{fig:metadata-viz} summarizes gender imbalance per genre (left) and t-SNE clusters (right).

We further embed the entire metadata into a 2D space using PCA-initialized 
t-SNE, as shown in Figure~\ref{fig:metadata-viz}. Genres form distinct clusters,
demonstrating that metadata features (genre, speed, gender, duration) carry 
meaningful separability, which can be exploited by downstream models.

To better understand the characteristics of the automatically generated QA pairs, 
we conducted a detailed statistical analysis of question types, answerable duration distributions, 
and their correlation with metadata attributes such as genre, speed, and gender.

\begin{figure}[t]
    \centering
    \includegraphics[width=0.95\linewidth]{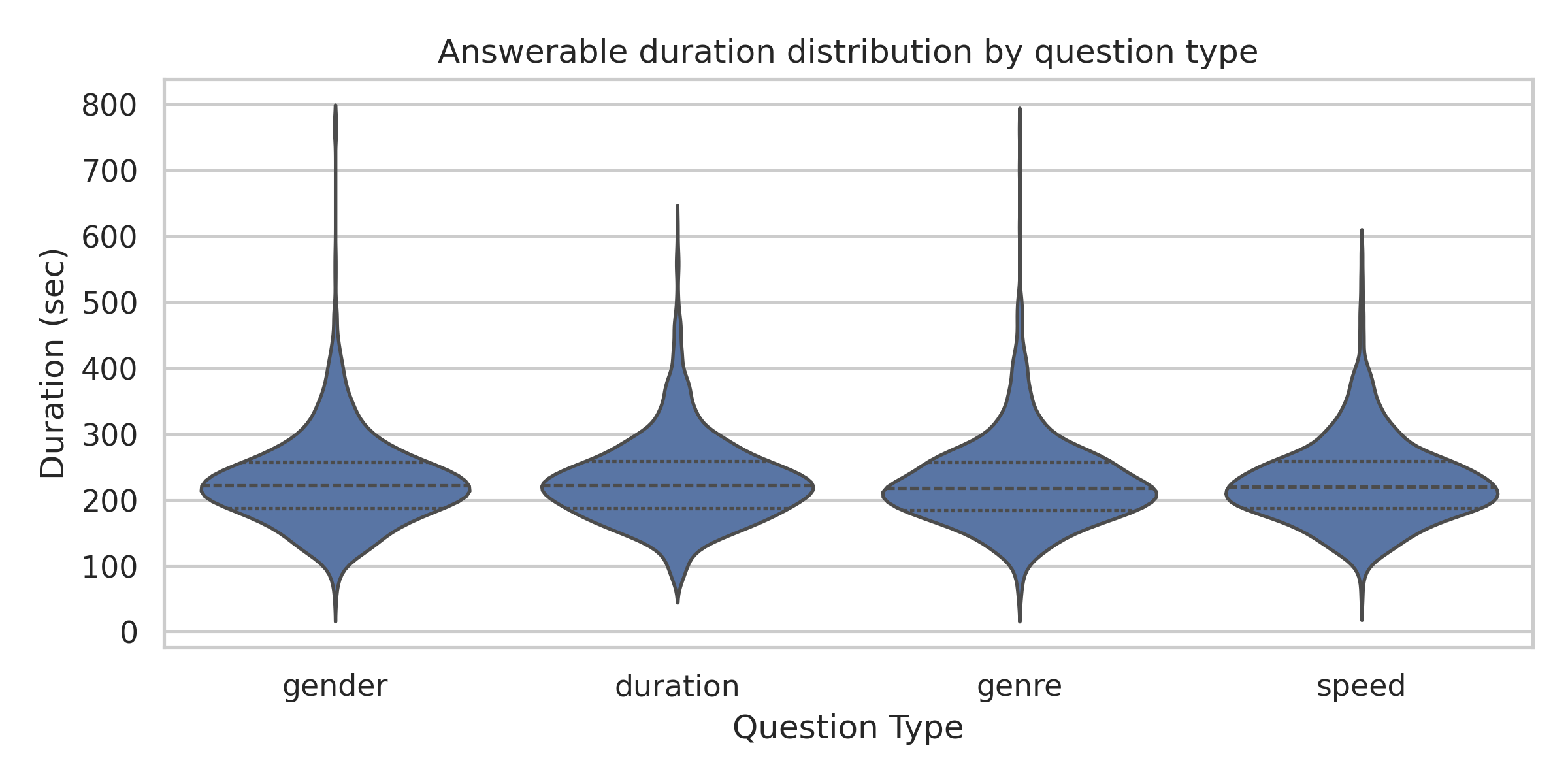}
    \caption{Distribution of track durations for each question type. 
    Most questions are asked about tracks between 180–260 seconds.}
    \label{fig:qa-duration}
\end{figure}

Jamendo-QA provides a balanced coverage across four major QA categories 
(\textit{genre}, \textit{speed}, \textit{duration}, and \textit{gender}), 
ensuring that models learn multiple dimensions of music understanding rather than overfitting to a single type. Across all question types, most tracks fall within 3–4 minutes of duration, consistent with popular-music structures, as shown in Figure~\ref{fig:qa-duration}.

\begin{figure}[t]
    \centering
    \includegraphics[width=0.95\linewidth]{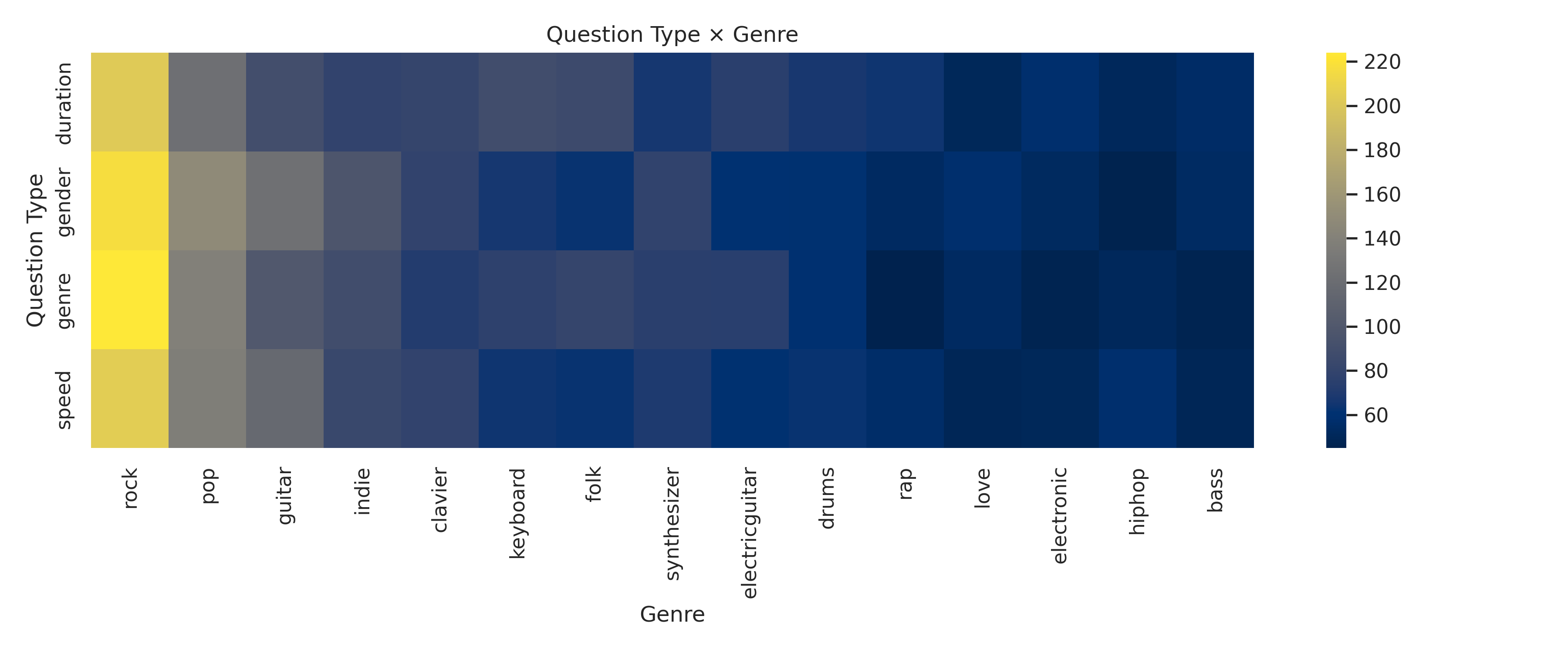}
    \caption{Question type frequency by genre (top-15 genres). 
    Darker colors indicate higher question density.}
    \label{fig:qa-genre-heatmap}
\end{figure}

Figure~\ref{fig:qa-genre-heatmap} highlights that \texttt{rock}, \texttt{pop}, and \texttt{guitar}-related tracks 
dominate all four question types. This reveals a potential bias toward rock/pop music, 
which should be considered when evaluating model generalization to underrepresented genres.

\begin{figure}[t]
    \centering
    \begin{minipage}[t]{0.48\linewidth}
        \centering
        \includegraphics[width=\linewidth]{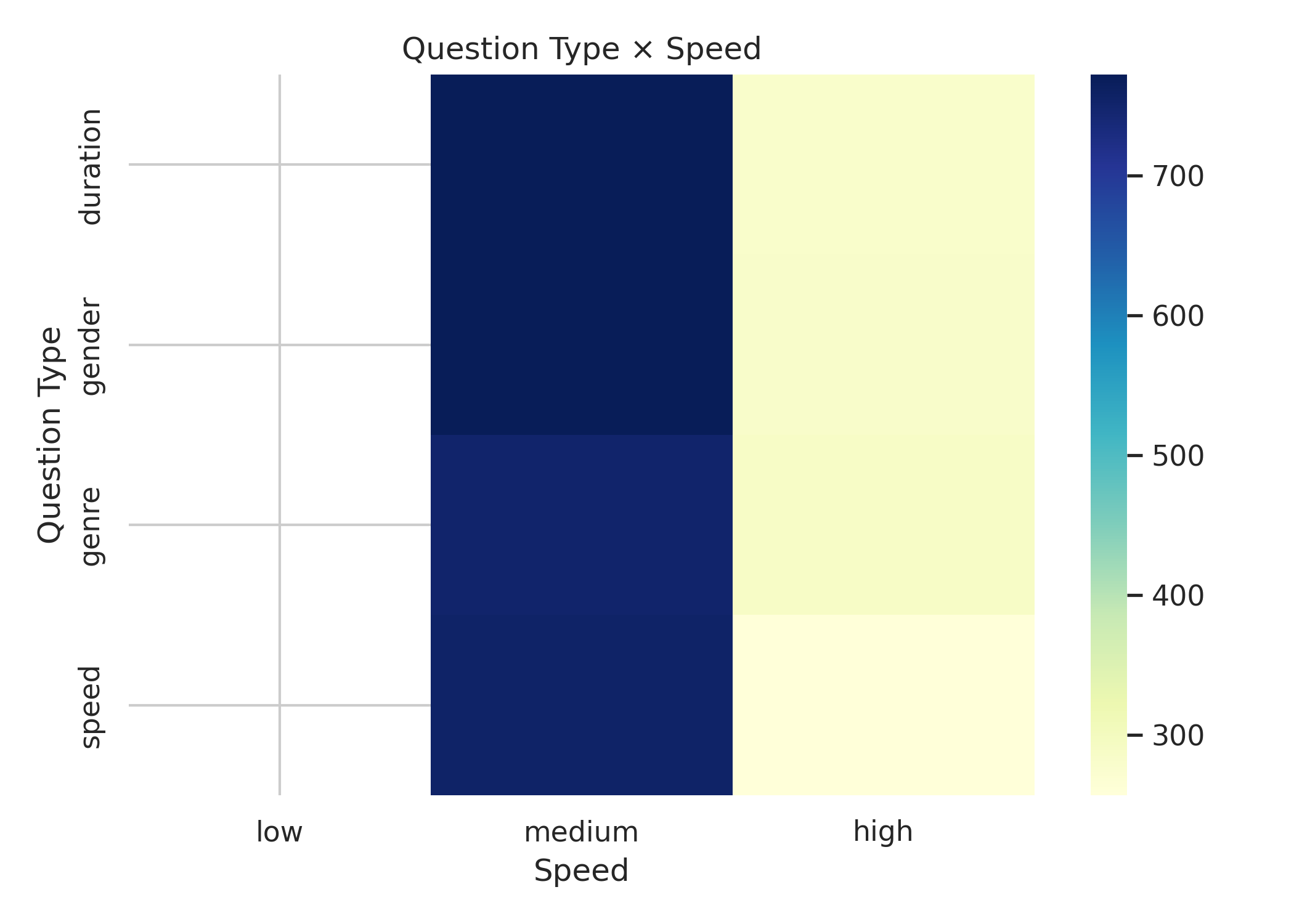}
    \end{minipage}\hfill
    \begin{minipage}[t]{0.48\linewidth}
        \centering
        \includegraphics[width=\linewidth]{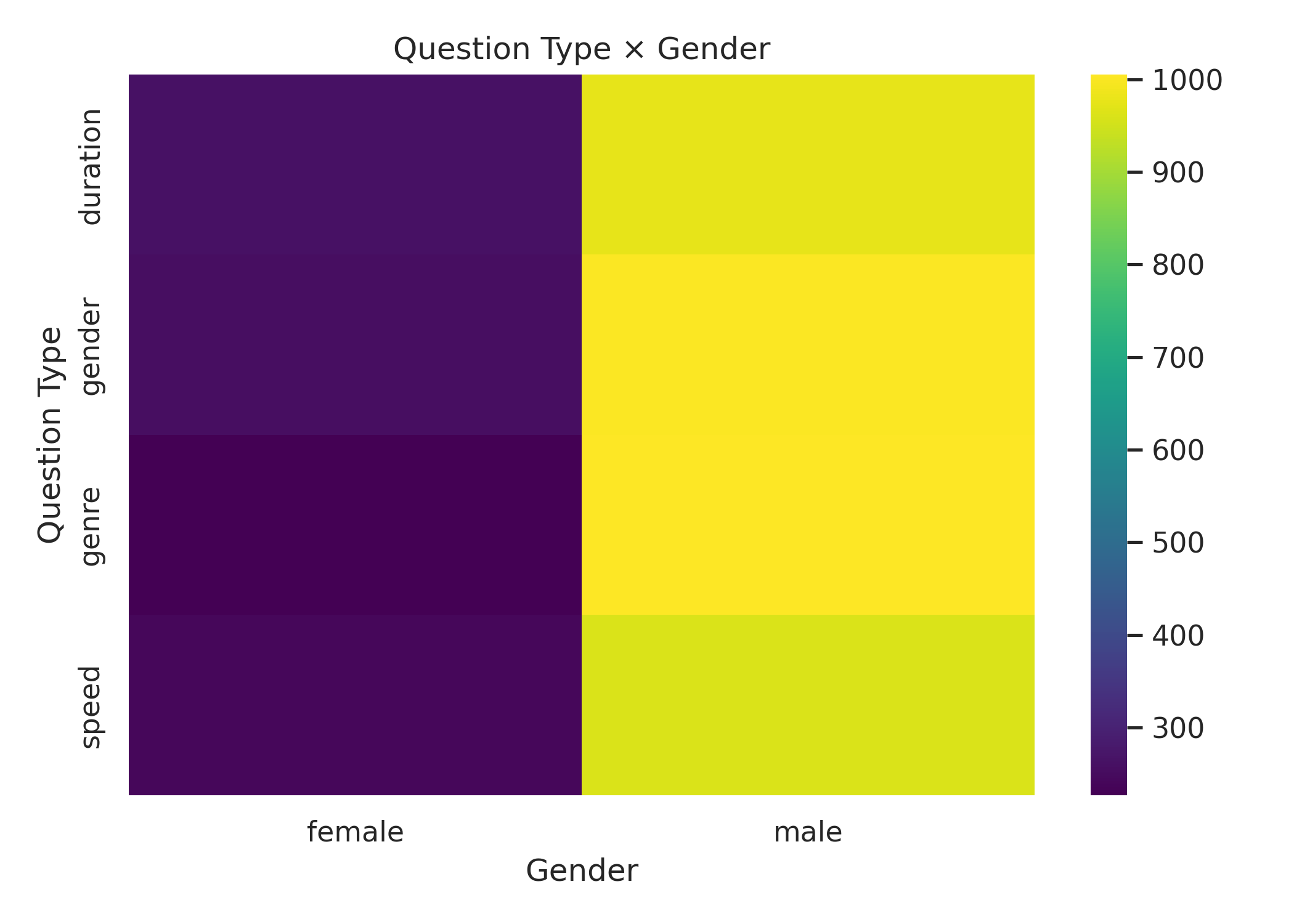}
    \end{minipage}
    \caption{QA distribution by metadata. Left: question type $\times$ speed (medium tempo dominates). Right: question type $\times$ gender (male-vocal skew).}
    \label{fig:qa-type-meta}
\end{figure}

We also observed that question–answer pairs are predominantly generated from \textit{medium}-tempo tracks and that male-vocal items are more frequent (Figure~\ref{fig:qa-type-meta}).

Finally, Figure~\ref{fig:qa-type-meta}(b) shows a strong skew towards male-vocal tracks across all question types.
This imbalance highlights the need for careful model evaluation to avoid gender bias in downstream applications.

\begin{table}[t]
\centering
\begin{threeparttable}
\caption{Comparison of metadata before and after imputation\tnote{a}}
\label{tab:metadata-example}
\setlength{\tabcolsep}{6pt}
\renewcommand{\arraystretch}{1.15}
\begin{tabularx}{\linewidth}{>{\raggedright\arraybackslash}X >{\raggedright\arraybackslash}X}
\toprule
\textbf{QA metadata} \\
\midrule
\begin{tabular}[t]{@{}l@{}}
\texttt{'audio\_path'}: `electronic\_ADreamWithinDream.wav',\\
\texttt{'genre'}: `electronic',\\
\texttt{'speed'}: `medium',\\
\texttt{'gender'}: `female',\\
\texttt{'length\_sec'}: 253,\\
\texttt{'lang'}: `en'
\end{tabular}
\\
\bottomrule
\end{tabularx}
\begin{tablenotes}
\footnotesize
\item[a] \textbf{Bold keys} represent the metadata fields, and their key–value pair will be used as a question–answer pair.
\end{tablenotes}
\end{threeparttable}
\end{table}

\subsection{Dataset Schema}
Each sample in Jamendo-QA contains an \texttt{audio\_path} pointing to the music file and a list of question–answer pairs under \texttt{conversations}. For a minimal example of the QA metadata structure,  Table~\ref{tab:metadata-example} provides a detailed overview of the dataset composition. It reports the number of samples per category and highlights the balance across different attributes. Full schema and data card are publicly available\footnote{\href{https://huggingface.co/datasets/m-a-a-p/Jamendo-QA}{HuggingFace: Jamendo-QA dataset}}.

\section{Conclusion}
\label{conclusion}

In this paper, we introduced Jamendo-QA dataset, a large-scale dataset for Music Question Answering (Music-QA). We addressed the significant challenge of data scarcity by leveraging the Qwen-Omni multimodal model to automatically generate high-quality question-answer pairs and captions from existing music audio. Our work is a pioneering effort in creating a dedicated, pure audio-based QA resource at scale, distinguishing it from existing datasets that are either cross-modal or rely on symbolic music representations.

Our analysis of the dataset reveals its key characteristics, including a typical distribution of track durations but a notable imbalance in genre tags, artist gender, and tempo. We believe these findings will be crucial for future work, encouraging researchers to develop more robust and fair models that can handle such data biases. However, Jamendo-QA provides a foundational resource for training and evaluating models for fine-grained musical reasoning, understanding, and retrieval. This dataset will foster new research directions in music information retrieval and generative AI, ultimately contributing to the development of more capable and versatile models for music understanding.

\newpage
\bibliographystyle{IEEEbib}
\label{sec:refs}
\bibliography{strings}

\begin{thebibliography}{10}

\bibitem{musicavqa}
Guangyao Li, Yake Wei, Yapeng Tian, Chenliang Xu, Ji-Rong Wen, and Di~Hu,
\newblock ``Learning to answer questions in dynamic audio-visual scenarios,''
\newblock {\em Proceedings of the IEEE/CVF Conference on Computer Vision and Pattern Recognition (CVPR)}, 2022.

\bibitem{mullama}
Shansong Liu, Atin~Sakkeer Hussain, Chenshuo Sun, and Ying Shan,
\newblock ``{Music Understanding LLaMA: Advancing Text-to-Music Generation with Question Answering and Captioning},''
\newblock {\em arXiv preprint arXiv:2308.11276}, 2023.

\bibitem{silu}
Stefan Elfwing, Eiji Uchibe, and Kenji Doya,
\newblock ``Sigmoid-weighted linear units for neural network function approximation in reinforcement learning,'' 2017.

\bibitem{mtgjamendo}
Dmitry Bogdanov, Minz Won, Philip Tovstogan, Alastair Porter, and Xavier Serra,
\newblock ``The mtg-jamendo dataset for automatic music tagging,''
\newblock in {\em Machine Learning for Music Discovery Workshop, International Conference on Machine Learning (ICML 2019)}, Long Beach, CA, United States, 2019.

\bibitem{jamendomaxcaps}
Abhinaba Roy, Renhang Liu, Tongyu Lu, and Dorien Herremans,
\newblock ``Jamendomaxcaps: A large scale music-caption dataset with imputed metadata,''
\newblock {\em arXiv:2502.07461}, 2025.

\bibitem{qwen2audio}
Yunfei Chu, Jin Xu, Qian Yang, Haojie Wei, Xipin Wei, Zhifang Guo, Yichong Leng, Yuanjun Lv, Jinzheng He, Junyang Lin, Chang Zhou, and Jingren Zhou,
\newblock ``Qwen2-audio technical report,'' 2024.

\bibitem{lpmusiccaps}
SeungHeon Doh, Keunwoo Choi, Jongpil Lee, and Juhan Nam,
\newblock ``Lp-musiccaps: Llm-based pseudo music captioning,'' 2023.

\bibitem{gpt3.5turbo}
Long Ouyang, Jeff Wu, Xu~Jiang, Diogo Almeida, Carroll~L. Wainwright, Pamela Mishkin, Chong Zhang, Sandhini Agarwal, Katarina Slama, Alex Ray, John Schulman, Jacob Hilton, Fraser Kelton, Luke Miller, Maddie Simens, Amanda Askell, Peter Welinder, Paul Christiano, Jan Leike, and Ryan Lowe,
\newblock ``Training language models to follow instructions with human feedback,'' 2022.

\bibitem{musicxqa}
Jian Chen, Wenye Ma, Penghang Liu, Wei Wang, Tengwei Song, Ming Li, Chenguang Wang, Jiayu Qin, Ruiyi Zhang, and Changyou Chen,
\newblock ``Musixqa: Advancing visual music understanding in multimodal large language models,'' 2025.

\bibitem{htsat}
Ke~Chen, Xingjian Du, Bilei Zhu, Zejun Ma, Taylor Berg-Kirkpatrick, and Shlomo Dubnov,
\newblock ``Hts-at: A hierarchical token-semantic audio transformer for sound classification and detection,'' 2022.

\bibitem{koizumi2020audiocaptioningusingpretrained}
Yuma Koizumi, Yasunori Ohishi, Daisuke Niizumi, Daiki Takeuchi, and Masahiro Yasuda,
\newblock ``Audio captioning using pre-trained large-scale language model guided by audio-based similar caption retrieval,'' 2020.

\bibitem{clotho}
Konstantinos Drossos, Samuel Lipping, and Tuomas Virtanen,
\newblock ``Clotho: An audio captioning dataset,'' 2019.

\bibitem{elizalde2023clap}
Benjamin Elizalde, Soham Deshmukh, Mahmoud~Al Ismail, and Huaming Wang,
\newblock ``Clap: Learning audio concepts from natural language supervision,'' 2022.

\bibitem{mert}
Yizhi LI, Ruibin Yuan, Ge~Zhang, Yinghao Ma, Xingran Chen, Hanzhi Yin, Chenghao Xiao, Chenghua Lin, Anton Ragni, Emmanouil Benetos, Norbert Gyenge, Roger Dannenberg, Ruibo Liu, Wenhu Chen, Gus Xia, Yemin Shi, Wenhao Huang, Zili Wang, Yike Guo, and Jie Fu,
\newblock ``{MERT}: Acoustic music understanding model with large-scale self-supervised training,''
\newblock in {\em The Twelfth International Conference on Learning Representations}, 2024.

\bibitem{mumullama}
Shansong Liu, Atin~Sakkeer Hussain, Qilong Wu, Chenshuo Sun, and Ying Shan,
\newblock ``Mumu-llama: Multi-modal music understanding and generation via large language models,''
\newblock {\em arXiv preprint arXiv:2412.06660}, 2024.

\bibitem{audioldm}
Haohe Liu, Zehua Chen, Yi~Yuan, Xinhao Mei, Xubo Liu, Danilo Mandic, Wenwu Wang, and Mark~D Plumbley,
\newblock ``{AudioLDM}: Text-to-audio generation with latent diffusion models,''
\newblock {\em Proceedings of the International Conference on Machine Learning}, pp. 21450--21474, 2023.

\bibitem{audioldm2}
Haohe Liu, Yi~Yuan, Xubo Liu, Xinhao Mei, Qiuqiang Kong, Qiao Tian, Yuping Wang, Wenwu Wang, Yuxuan Wang, and Mark~D. Plumbley,
\newblock ``Audioldm 2: Learning holistic audio generation with self-supervised pretraining,''
\newblock {\em IEEE/ACM Transactions on Audio, Speech, and Language Processing}, vol. 32, pp. 2871--2883, 2024.

\bibitem{musicgen}
Jade Copet, Felix Kreuk, Itai Gat, Tal Remez, David Kant, Gabriel Synnaeve, Yossi Adi, and Alexandre Défossez,
\newblock ``Simple and controllable music generation,'' 2024.

\bibitem{lstm}
Sepp Hochreiter and J\"{u}rgen Schmidhuber,
\newblock ``Long short-term memory,''
\newblock {\em Neural Comput.}, vol. 9, no. 8, pp. 1735–1780, Nov. 1997.

\bibitem{qwenomni}
Jin Xu, Zhifang Guo, Jinzheng He, Hangrui Hu, Ting He, Shuai Bai, Keqin Chen, Jialin Wang, Yang Fan, Kai Dang, Bin Zhang, Xiong Wang, Yunfei Chu, and Junyang Lin,
\newblock ``Qwen2.5-omni technical report,'' 2025.

\end{thebibliography}

\end{document}